\documentclass{article}

\font\sqi=cmssq8
\def\DR{\rm I\kern-1.45pt\rm R}
\def\DC{\kern2pt {\hbox{\sqi I}}\kern-4.2pt\rm C}
\newcommand{\ben}{\begin{enumerate}}
\newcommand{\een}{\end{enumerate}}
\newcommand{\beq}{\begin{equation}}
\newcommand{\eeq}{\end{equation}}
\newcommand{\bse}{\begin{subequation}}
\newcommand{\ese}{\end{subequation}}
\newcommand{\bea}{\begin{eqnarray}}
\newcommand{\eea}{\end{eqnarray}}
\newcommand{\bc}{\begin{center}}
\newcommand{\ec}{\end{center}}

\newcommand{\bs}{\mbox{\boldmath $\sigma$}}
\def\r{r_0}
\newcommand{\ch}{{\tt h}}

\def\DH{\rm I\kern-1.5pt\rm H\kern-1.5pt\rm I}
\begin{document}
%\hfill{hep-th/9911020}
%\vspace{0.6cm}
\begin{center}
{\large\bf  (Super)Oscillator on $\DC P^N$ and Constant Magnetic Field} \\
\vspace{0.5 cm}
{\large Stefano Bellucci$^1$ and   Armen Nersessian$^{2,3}$ }
\end{center}
{\it $^1$ INFN-Laboratori Nazionali di Frascati,
 P.O. Box 13, I-00044, Italy\\
$^2$ Yerevan State University, Alex  Manoogian St., 1, Yerevan,
375025, Armenia\\
$^3$ Yerevan Physics Institute, Alikhanian Brothers St., 2, Yerevan, 375036,
 Armenia}
\begin{abstract}
We define the ``maximally integrable" isotropic oscillator on  $\DC P^N$
 and discuss its various properties, in particular, the behaviour
of the system with respect to a constant magnetic field.
We show that the properties of the oscillator on $\DC P^N$  qualitatively
 differ in the $N>1$ and $N=1$ cases.
  In the former case we construct the ``axially symmetric'' system
which is  locally equivalent to the oscillator.
We perform the Kustaanheimo-Stiefel transformation  of the oscillator on
 $\DC P^2$ and construct some generalized MIC-Kepler problem.
We also define a ${\cal N}=2$ superextension
of the oscillator on $\DC P^N$ and show that
for $N>1$ the inclusion of a constant magnetic field
 preserves the supersymmetry of the system.
\end{abstract}
\begin{center}
\end{center}
\setcounter{equation}0
\section{Introduction}
The harmonic oscillator plays a distinguished role in theoretical and
mathematical physics, due to its overcomplete symmetry group. The wide number of hidden
symmetries provides the oscillator with unique properties, e.g. closed
classical trajectories, the
 degeneracy of the quantum-mechanical
 energy spectrum, the separability of variables in a few
coordinate systems.
The overcomplete symmetry
allows one to preserve the exact
 solvability of the oscillator, even after some
deformation of the potential
breaking the initial symmetry of the system.
 Particulary, the oscillator remains exactly solvable
 after coupling to a constant magnetic field, though the latter  removes
the hidden symmetries of the system.
The reduction of the oscillator  to low dimensions allows one to
construct new
integrable systems  with hidden symmetries
(in fact, almost all  integrable systems of classical and quantum
mechanics
are related with
either the free particle case, or the oscillator) \cite{perelomov}.
There is a nontrivial  relation between oscillator and
 Coulomb systems:
the $(N+1)-$dimensional  Coulomb problem can be obtained from the
 $2N-$dimensional oscillator
by the so-called   Levi-Civita (or Bohlin), Kustaanheimo-Stiefel and
Hurwitz transformations, when  $N=1, 2,4$ \cite{bohlin}.
The transformations  correspond to the reduction of the oscillator
 by the actions of
$Z_2$, $U(1)$ and $SU(2)$ groups, respectively,
and are  based on the Hopf maps $S^1/Z_2=S^1$,
 $S^3/U(1)=\DC P^1\cong S^2$,
$S^7/SU(2)=\DH P^1\cong S^4$ (relating the
the angular parts of the oscillator and Coulomb  problems).
Indeed, reducing the oscillators
 we get  some parametric families  of Coulomb-like systems,
specified by the presence of a magnetic flux  for $N=1$; by
a Dirac monopole  for $N=2$ (the MIC-Kepler system); and by a
Yang monopole\footnote{Under ``Yang monopole'' we mean a five-dimensional
$SU(2)$ generalization of a Dirac monopole \cite{yang}.}  for $N=4$
(see, respectively, \cite{1,2,3}).
It could be checked easily, that the MIC-Kepler  system, initially introduced by
Zwanziger for the
description of the
relative motion of two Dirac  dyons,  also describes the
  scattering of two well-separated BPS monopoles and dyons.
   The latter problem was considered
in a well-known paper  by Gibbons and Manton \cite{gm},
 where the existence of a hidden
Coulomb-like symmetry was established
(see also \cite{ho1}). Let us
mention also  the key role of the
Hurwitz transformation
(and of the second Hopf map) in the recently proposed
 higher-dimensional quantum Hall effect \cite{Zhang:2001xs}
(see also \cite{kn,ber}).

The oscillator is a distinguished system, also with  respect
to supersymmetrization.
A supersymmetric oscillator is specified
 by the splitting of fermionic and bosonic degrees of freedom. Thus,
it inherits the hidden symmetries of the initial system.
 We notice that the construction of integrable supersymmetric
 mechanics is interesting
 not only in a field-theoretical context.
Being in deep connection with the factorization problem, the
supersymmetrization of integrable systems
could yield a new set of integrable systems with isospectral
potentials.
 Since the list of references  on supersymmetric mechanics is enormous,
we  refer to  the introductory reviews \cite{sukh}
 (mostly devoted to the  connection
of supersymmetric quantum mechanics with the factorization problem) and
\cite{lima} (containing  the most  complete list of  references
on field-theoretical aspects of supersymmetric mechanics).

Recent progress in string theory inspired  interest for noncommutative
field theories \cite{douglas}
 and, in particular,
for noncommutative quantum mechanics \cite{ncsph}.
 The oscillator  was found
to be a distinguished case in noncommutative quantum mechanics too:
at the moment it is the only exactly solved (even in the presence
of a constant magnetic field) noncommutative quantum mechanical system
with a non-zero potential \cite{ncpl}.

There is nontrivial generalization of the oscillator
on the sphere and the two-sheet
hyperboloid (pseudosphere) \cite{higgs} given  by the
  potential
\begin{equation}
U_{osc}=\frac{\omega^2 \r^2}{2}\frac{{\bf x}^2}{{x}^2_{d+1}}.
\end{equation}
Here ${\bf x}, x_{d+1}$ are the (pseudo)Euclidean coordinates
of the ambient
space $\DR^{d+1}$($\DR^{d.1}$):
$\epsilon{\bf x}^2+ x^2_{d+1}=\r^2$, with $\epsilon=+1$ for  the sphere,
$\epsilon=-1$ for the pseudosphere.\\
This system has
 a nonlinear hidden symmetry algebra, providing it
  with properties similar to those
 of a conventional oscillator.
Applying to the oscillator on the (pseudo)sphere the standard
 Levi-Civita, Kustaanheimo-Stiefel and Hurwitz transformations,
one can obtain the generalization of
flux-Coulomb, MIC-Kepler and Yang-Coulomb systems on the
(pseudo)sphere \cite{np}.\footnote{Let us remind, that the Coulomb system
on the (pseudo)sphere is defined by the potential
\cite{sch}
$$
U_{C}=-\frac{\gamma}{\r}\frac{x_{d+1}}{|{\bf x}|}.$$
Quantum mechanics of the oscillator
and Coulomb system on the $D-$dimensional
sphere and pseudosphere is considered in detail
in Ref. \cite{george}.}
In the present
 paper we define the  oscillator on
complex
 projective spaces
$\DC P^N$, from the requirement that it
possesses  hidden symmetries  generalizing those
of the planar oscillator,
and consider its behaviour with respect to the coupling
to a constant magnetic field.

The oscillator on $\DC P^1=S^2$ coincides
with the Higgs oscillator
on the sphere $S^2$ (note that $\DC P^1= S^2$).
The oscillator on $\DC P^N$, $N>1$ is defined
 by the potential
 \beq
\begin{array}{c}
U(z\bar z)=\omega^2\r^2 z\bar z,
\end{array}
\eeq
where $z^a, \bar z^a$ are inhomogeneous coordinates
of $\DC P^N$, corresponding to the
Fubini-Study metric
\beq
g_{a\bar b}dz^ad\bar z^b=\r^2\frac{dzd\bar z}{1+z\bar z}-
\r^2\frac{(\bar z dz)(zd\bar z)}{(1+z\bar z)^2}.
\eeq
In contrast  to the case of the
oscillator on $\DC P^1 = S^2$ which is
 defined  on the disk $|z|<1$, the oscillator  on $\DC P^N$, $N>1$
is defined on the whole chart.
The transition to another
 chart of  $\DC P^N$ transforms the oscillator into the system with the
potential
$$
U= \omega^2\r^2\left(
\frac{1}{z^1\bar z^1}+
\frac{z^2{\bar z}^2+\ldots + z^N {\bar z}^N}{z^1\bar z^1}\right),
$$
which has the oscillator  symmetry algebra.

The Kustaanheimo-Stiefel transformation of the oscillator on $\DC P^2$
yields a generalization
 of the MIC-Kepler system, which can be transformed
into the  MIC-Kepler system on the three-dimensional  hyperboloid.

 The oscillator on $\DC P^N$ admits,
because of its  K\"ahler structure,
 a simple  coupling to a constant
 magnetic field. This can be achieved by carrying out the following replacement of
the symplectic structure:
 $\Omega_0\to\Omega_0+iBg_{a\bar b}dz^a\wedge d\bar z^b$.
The coupling to a constant magnetic field preserves all symmetries of the oscillator,
including kinematical and hidden ones, when $N>1$, but kinematical symmetries only, when $N=1$.

Below, we construct the ${\cal N}=2$
supersymmetric oscillator on $\DC P^N$ and study its
behaviour, with respect to the coupling to a constant magnetic  field
(the oscillator on $\DC P^N$, in contrast with the one
on $\DC^N$, does not admit the ${\cal N}=4$ supersymmetrization).
We  show that, in contrast with the ${\cal N}=2$
superoscillator on $\DC P^1=S^2$,
 the  ${\cal N}=2$ superoscillator on  $\DC P^N$, $N>1$
allows  coupling to a  constant magnetic field,
without breaking supersymmetry.

\setcounter{equation}0
\section{Oscillator on $\DC P^N$}

This section is devoted to the construction of the
oscillator system on the complex projective space
$\DC P^N$. Our consideration essentially exploits the fact
 that the complex projective space is a
constant curvature  K\"ahler manifold.
Hence, our model  could be easily adopted for the formulation
 of the oscillator system  on the other spaces of that sort.

Let us remind that the K\"ahler manifold $M$
is  equipped with the metric, which could be
 locally represented in the form
  \begin{equation}
  g_{a \bar b}dz^ad{\bar z}^b =
\frac{\partial^2 K}{\partial z^a \partial{\bar z}^{b}}dz^ad{\bar z}^b,
  \end{equation}
and with the associated Poisson bracket
\begin{equation}
   \{ f,g\}_0 =
 i\frac{\partial  f}{\partial {\bar z}^a}g^{{\bar a}b}
\frac{\partial  g}{\partial z^b} -
i\frac{\partial g}{\partial z^b}g^{{\bar a}b}
\frac{\partial  f }{\partial {\bar z}^a},\quad
g^{{\bar a}b}
g_{b{\bar c}}=\delta^{\bar a}_{\bar c}.      \label{p0}
\end{equation}
The local real function $K(z,\bar z)$
 is called the K\"ahler potential.

The complex projective space  $\DC P^N$
could be  equipped with the Fubini-Study metric,
 given by the K\"ahler potential
  \begin{equation}
  K=\r^2\log (1+ z\bar z).
  \end{equation}
The scalar curvature of $\DC P^N$ is related with
the parameter $\r^2$
 as follows: $R=N(N+1)/\r^2$.\\
The isometries of the K\"ahler structure
are generated by
the {\it holomorphic  Hamiltonian vector fields}
 \begin{equation}
{\bf V}_{\mu}=
    V_\mu^{a}(z)\frac
{\partial}{\partial z^a}+
{\bar V}_\mu^{\bar a}(\bar z) \frac{\partial}
{\partial \bar z^a}, \quad  [{\bf V}_{\mu},{\bf V}_{\nu}]=
C_{\mu \nu}^{\lambda}{\bf V}_{\lambda} ,
\end{equation}
where
\begin{equation}
{\bf V}_\mu=\{\ch_\mu, \}_0 , \quad
 \{ \ch_{\mu},\ch_{\nu}\}_0=
C_{\mu \nu}^{\lambda}
\ch_{\lambda},\quad
\frac{\partial^2 \ch_\mu}{\partial z^a \partial z^b} -
\Gamma^c_{ab}\frac{\partial \ch_\mu}{\partial z^c}=0.
\label{v}\end{equation}
The real functions $\ch_\mu$ are called Killing potentials.

The symmetry algebra of  $\DC P^N $  is $su(N+1)$.
This algebra is defined  by the Killing  potentials
\beq
\ch_T= T^{\bar a b}\ch_{\bar a b}- {\rm tr}\;{\hat T}, \quad
\ch^1_a=\ch^-_a+\ch^+_a,\quad \ch^2_a=i(\ch^-_a-\ch^+_a),
\eeq
where
\begin{equation}
\ch_{\bar a b}=\r^2\frac{z^a \bar z^b}{1+ z\bar z},\quad
\ch^-_a=\r^2\frac{z^a}{1+z\bar z},\quad
\ch^+_a=\r^2\frac{\bar z^a}{1+ z\bar z},
\end{equation}
and ${\hat T}$ are $N\times N$ Hermitean matrices:
${T}^{\bar a b}= {\bar{T^{\bar b a}}}$.

The algebra of $\ch_{\bar a b}, \ch^\pm_a$ reads:
\beq
\begin{array}{c}
\{\ch_{{\bar a} b}, \ch_{\bar c d}\}_0=
i\delta_{\bar a d}\ch_{\bar b c}
-i\delta_{\bar c b}\ch_{\bar a d},\\
\{\ch^-_a, \ch^+_b\}_0=
i\delta_{\bar a b}(\r^2-
{\rm tr}\;\ch_{\bar a b})+i\ch_{\bar a b},
\quad\{\ch^{\pm}_a, \ch^{\pm}_b\}_0=0,
\quad
\{\ch^{\pm}_a, \ch_{\bar b c}\}_0=
\mp i\ch^{\pm}_b\delta_{ a b}\quad .
\end{array}
\eeq

Let us equip  the cotangent bundle $T_*\DC P^N$
with the  symplectic structure
\begin{equation}
\Omega_B=
dz^a\wedge d\pi_a +
d{\bar z}^{a}\wedge d{\bar\pi}_{a} +iBg_{a\bar b}dz^a\wedge
d{\bar z}^b,
\label{ssB}\end{equation}
which defines, together with  the Hamiltonian
\begin{equation}
  {\cal D}= g^{a \bar b}\pi_a{\bar \pi}_b \quad ,
\label{do}\end{equation}
 the dynamics of a free particle on
$\DC P^N$, in the presence of a constant
magnetic field $B$.
The isometries of a K\"ahler
 structure define the
Noether's constants of motion of a free particle
 \begin{equation}
{\cal J}_{\mu}\equiv J_\mu+ B\ch_\mu=V_\mu^{a}\pi_a +
 {\bar V}_{\mu}^{\bar a} {\bar\pi}_{\bar a} +B\ch_\mu :
\left\{
\begin{array}{c}
  \{{\cal D}, J_{\mu}\}=0, \\
  \{J_\mu, J_\nu\}=C_{\mu\nu}^\lambda J_\lambda .
\end{array}\right.\label{jmu}\end{equation}
Explicitly, we have
  \begin{equation}
\left.
\begin{array}{c}
J_{a\bar b}=-{i}z^b\pi_a+ i\bar\pi_b \bar z^a \;,\quad
iJ^{+}_a=\pi_a+\bar z^a(\bar z\bar\pi),\quad
-iJ^{-}_a=\bar\pi_a+ z^a(z\pi) .
\end{array}\right.
\label{jab}\end{equation}
Notice that the vector fields
 generated by ${\cal J}_\mu$ are independent on $B$
\begin{equation}
{\bf{\tilde V}}=V^a(z)\frac{\partial}{\partial z^a}-V^a_{,b}\pi_a
\frac{\partial}{\partial \pi_a}+
{\bar V}^a(\bar z)\frac{\partial}{\partial\bar z^a}
-{\bar V}^a_{,\bar b}\bar\pi_a
\frac{\partial}{\partial \bar\pi_a}\quad.
\end{equation}
Hence, the inclusion of a
 constant magnetic field preserves
the whole symmetry algebra of a free
particle moving in a K\"ahler space. \\

Now, let us   consider  the $u(N)$-invariant  Hamiltonian
\begin{equation}
 {\cal H}=g^{a{\bar b}}\pi_a{\bar \pi}_b +U(z\bar z),
\label{du}\end{equation}
and require it to have the hidden
symmetry (similar to the one of the oscillator)
given by either one of the constants of motion
\begin{equation}
\begin{array}{cc}
i)\quad &
I^+_{a b}=
{\cal J}^+_a {\cal J}^+_{ b} +
 f_+(z\bar z) {\bar z}^a  {\bar z}^b,\\
ii)\quad
 &I_{  a \bar b}={\cal J}^+_a {\cal J}^-_{\bar b}
 + f_0(z\bar z) {\bar z}^a {z}^{ b}.
\end{array}
\label{ip}\end{equation}
 Straightforward calculations immediately yield the following
constraints:
\beq
 \begin{array}{ccccc}i)\quad &
 B=0&N=1\;&
U(x)=c_1 x/({1-x})^2+c_0& f_+=c_1/{(1-x)^2},\\
ii)\quad &
{\rm for\; any\; value\; of\;} B\;&N=1,2\ldots &U(x)=c_1x+c_0& f_0=c_1 .
\end{array}
\eeq
Taking into account that
 ${\cal H}= {\rm Tr}\; {\hat I}+ { \rm Tr}\; {\hat J}^2/2\r^2 $,
we get the
 following generalizations of the oscillator on $\DC P^N$.
\begin{itemize}
\item   $\DC P^1$.
The  oscillator is defined by the Hamiltonian system
\beq
{\cal  H}=
\frac{(1+z\bar z)^2 {\pi{\bar\pi}}}{\r^2}
+\frac{\omega^2\r^2{z\bar z}}{(1-{z\bar z})^2},
\quad \Omega_0=dz\wedge d\pi+d\bar z
\wedge d{\bar\pi}.
\label{ho}\end{equation}
The   symmetry  algebra is given by the $U(1)$
generator $J$
and  the complex (or vectorial) constant of motion  $I^\pm$
\begin{equation}
J=i(\pi z-\bar\pi \bar z),\;\; { I}_+=\frac{{{ J}^2_+}}{\r^2} -
\frac{\omega^2\r^2{\bar z}^2}{(1- z\bar z)^2}\;:\;
 \{J, { I}_\pm\}=\pm 2i{ I_\pm},\;\;\{{ I}_-,{ I}_+\}=4i
\left(\omega^2 J +\frac{J{\cal H}}{\r^2}-
\frac{J^3}{2\r^4}\right).%\\
\label{Ia}\end{equation}
This is nothing  but the well-known Higgs oscillator
on the sphere $S^2=\DC P^1$ \cite{higgs}.
\item $\DC P^N$, $N>1$.
The oscillator is defined by the Hamiltonian system
\beq
\begin{array}{c}
{\cal H}=g^{a\bar b}\pi_a\bar\pi_b +
\omega^2\r^2 z\bar z,\quad
\Omega_0=
dz^a\wedge d\pi_a +d{\bar z}^{a}\wedge d{\bar\pi}_a .
\end{array}\label{hn}\eeq
Its symmetries  are given by the  constants of motion
\beq
J_{a\bar b}={i}(z^b\pi_a-\bar\pi_b\bar z^a), \quad
I_{a\bar b}=
\frac{J^+_a J^-_b}{\r^2} +\omega^2\r^2 {\bar z}^a z^b\; ,
\label{sym}\eeq
which define the   nonlinear (quadratic) algebra
\beq
\begin{array}{c}
\{J_{{\bar a} b}, J_{\bar c d}\}=
i\delta_{\bar a d}J_{\bar b c}
-i\delta_{\bar c b}J_{\bar a d},\quad
\{I_{a\bar b}, J_{c\bar d}\}=
i\delta_{c\bar b}I_{a\bar d}-i\delta_{a\bar d}I_{c\bar b} \\
 \{I_{a \bar b}, I_{c\bar d}\}=
i\omega^2 \delta_{c\bar b} J_{a\bar d}- i\omega^2
 \delta_{a\bar d}J_{c\bar b}

+i I_{c\bar b}(J_{a\bar d}+J_0\delta_{a\bar d})/\r^2
- i I_{a\bar d}(J_{c\bar b}+J_0\delta_{c\bar b})/\r^2 \quad .
\end{array}
\label{cpnalg}\eeq

\end{itemize}
It is convenient to  introduce
the generators
\beq
J_i=T_i^{a\bar b}J_{a\bar b},\quad
 J_0={\rm Tr} \;{\hat J},
\quad I_i=T_i^{a\bar b}I_{a\bar b},\quad
I_0={\rm Tr}\; {\hat I}\;,
\eeq
where $T_i$ are traceless  $N\times N$ Hermitean matrices
(the generators of
the  $su(N)$ algebra). The above generators belonging
to the center of algebra read:
 \beq
J_0=i(z\pi-\bar\pi\bar z),\quad
{\cal H}_{N>1}=
 I_0+ \frac{{ \rm Tr}\; {\hat J}^2 +J^2_0}{2\r^2} \quad .
\eeq
Also the following equality holds
\beq
{ \rm Tr }{\hat I}^2+
\omega^2{ \rm Tr }{\hat J}^2=I_0^2+\omega^2J^2_0 \quad .
\eeq
We have got the
 ``maximally integrable"  generalization of
the oscillator on complex
 projective spaces, i.e.  the system
with the highest possible
 number of functionally independent constants of motion.\footnote{
In the theory of integrable systems such systems  are  called
``maximally superintegrable systems". We prefer to suppress
the prefix ``super" in this context,
in order to avoid any confusion with supersymmetric systems.}

We established the following essential properties of the latter
system.
\begin{itemize}
\item The oscillator on  $\DC P^N$, $N>1$ is well-defined on the
whole chart of the
 complex projective space, $0<|z|<\infty$. The oscillator
on $\DC P^1 \sim S^2$ (as well as on higher-dimensional spheres)
is defined on the disc $|z|<1$ only.
 The constant magnetic field preserves the
hidden symmetries of the oscillator on $\DC P^N$, for $N>1$.

\item The above construction could be easily extended
 for the noncompact version of $\DC P^N$,
provided by  the Lobachewski space
 ${\cal L}^N=SU(1.N)/U(1)\times SU(N)$.
For this purpose, we should  replace the Fubini-Study metric
 with the one generated by the  K\"ahler
potential $K=-\r^2\log(1-z\bar z)$,
  and subsequently replace the
   Killing potentials and Noether constants of
 $\DC P^N$ with the ones of ${\cal L}^N$.
The Killing potentials of ${\cal L}^N$ are
 defined by the functions
\begin{equation}
\ch_{\bar a b}=-\r^2\frac{z^a \bar z^b}{1-z\bar z},\quad
\ch^-_a=-\r^2\frac{z^a}{1-z\bar z},\quad
\ch^+_a=-\r^2\frac{\bar z^a}{1- z\bar z}\quad.
\end{equation}
 \end{itemize}

Globally, the complex projective space $\DC P^N$
is covered by $N+1$ charts, marked by the
indices ${\tilde a}=0,a$. The transition functions
from the ${\tilde b}$-th chart to
the ${\tilde c}$-th one are of the form
\beq
z^{\tilde a}_{({\tilde c})}=
\frac{z^{\tilde a}_{({\tilde b})}}{z^{\tilde c}_{({\tilde b})}},
\quad {\rm where}\quad z^{\tilde a}_{({\tilde a})}=1 .
\label{trans}\eeq
On $\DC P^1$ the transition functions take the
simple form $z\to 1/z$,
corresponding to the transition
from one hemisphere
to the other.
 The respective transformation of the
 momenta is $\pi\to -z^2\pi$.
The Hamiltonian of the oscillator on $\DC P^1$ is
obviously invariant under the
above transformation.
In higher-dimensions we get a rather different picture,
since the potential term is not covariant under the
transition (\ref{trans}).
Let us consider this transformation in more details.

 The transition functions (\ref{trans})
define the following canonical transformation,
 which is singular on the $z^1=0$ ``axes":
\beq
z^1\to 1/z^1,\quad\pi_1\to -z^1(z\pi), \quad
z^{\hat a}\to z^{\hat a}/z^1,\quad
 \pi_{\hat a}\to z^1\pi_{\hat a}
\quad
{\hat a}=2,\ldots N.
\label{tz}\eeq
The kinetic term   is covariant with respect
to the above transformation,
while the potential term is not.
As a result, we get the integrable system
on $\DC P^N$, $N>1$ defined
by the Hamiltonian
\beq
{\cal H}_{\rm Back}=g^{a\bar b}\pi_a \bar\pi_b+
\omega^2\r^2\left(
\frac{1}{z^1\bar z^1}+
\frac{z^2{\bar z}^2+\ldots +
 z^N {\bar z}^N}{z^1\bar z^1}\right).
\label{hb}\eeq
This  system inherits the whole symmetry
 algebra  of the oscillator,
 i.e.  it is  a ``maximally integrable" system.
Its constants of motion can be obtained by a
 straightforward transformation of those of the
oscillator, given in
(\ref{sym}).
Note that, in spite of its
``maximal integrability", the system is
 not invariant under ``spatial" $u(N)$ rotations.

On the Lobachewski space ${\cal L}^N$, $N>1$
there is  no analog of this system.
The ``ambient" space  for the Lobachewski
 plane is $\DC^{1.N}$. The transitions (\ref{trans}) transform
the oscillator  on ${\cal L}^N$ into
a system on
the space with the signature $(-,-,+, \ldots,+)$.

\subsection*{$\DC P^2$: Kustaanheimo-Stiefel transformation}
As we mentioned in the Introduction, the oscillator
on two-, four-, and eight- dimensional  planes
and spheres could be reduced
 to the two- , three- and five- dimensional Coulomb systems,
  and their generalizations specified by the
presence of  monopoles.
Particularly,  the oscillator on
$S^2=\DC P^1$  and $AdS_2={\cal L}$ can
be reduced, by the so-called
Levi-Civita transformation,
to the  Coulomb systems on two-dimensional
hyperboloid (Lobachewski plane)
 ${\cal L}$.
Similarly, the Kustaanheimo-Stiefel transformation
of the  oscillator on
a four-dimensional sphere and a
four-dimensional two-sheet hyperboloid
leads to  the generalization of
 the MIC-Kepler problem
on a three-dimensional two-sheet
hyperboloid  \cite{np}.

Let us    consider the behavior  of the oscillator on
$\DC P^2$, with respect to  the
Kustaanheimo-Stiefel transformation.
The constants of motion  of the
oscillator on $\DC P^2 $
are given by the generators
\beq
{\bf I}=\frac{J_+{\bs}J_-}{\r^2} +
\omega^2\r^2 z{\bs}\bar z,\quad
{\bf J}=iz{\bs}\pi-i\bar\pi{\bs}\bar z,
\quad J_0=iz\pi-i\bar z\bar\pi,
\eeq
where $\bs$ denotes  standard Pauli matrices.\\
Their algebra  reads:
\beq
\begin{array}{c}
\{J_0, {I}_k\}=\{J_0, {J}_k\}=0,\quad
\{J_k,J_l \}=2\epsilon_{klm}J_m,\quad
 \{I_k, J_l \}=2\epsilon_{klm}I_m,\\
\{I_k,I_l \}=
\epsilon_{klm}\left(2\omega^2 J_m -3I_m J_0/\r^2
+I_0 J_m/\r^2\right).
\end{array}
\label{micalg}\eeq
In order to reduce this  system by the
 Hamiltonian action of  $J_0$, we have
to fix its  value
\beq
J_0=2s,
\eeq
and then factorize the level surface by the $U(1)$ group action.
The resulting six-dimensional phase space $T^*M^{\rm red}$
can be parameterized by the following  $U(1)$-invariant
 functions:
\beq
 {\bf x}=z\bs{\bar z},\quad
 {\bf p}=\frac{z\bs\pi +\bar\pi\bs\bar z}{2z\bar z}\;
:\quad \{{\bf x}, J_0\}=\{{\bf p}, J_0\}=0.
\end{equation}
In these coordinates
 the reduced  symplectic structure and the generators of the
 angular momentum are given by the expressions
\begin{equation}
\Omega_{\rm red}=d{\bf p}\wedge d{\bf x} +
s\frac{{\bf x}\times d{\bf x }\times d{\bf x}}{|{\bf x}|^3},
\quad
{\bf J}_{red}={\bf J}/2=
{\bf p}\times{\bf x} + s\frac{{\bf x}}{ |{\bf x}|  }.
\label{ss2}\end{equation}
Thus, the reduced system is specified by the presence
of a Dirac  monopole.

The reduced Hamiltonian is given by the expression
\begin{equation}
 {\cal H}_{red}=\frac{(1+x)}{\r^2}\left[x{\bf p}^2 +
 {({\bf x} {\bf p})^2}\right]
+ s^2\frac{(1+x)^2}{\r^2x} +\omega^2\r^2 x,
 \quad {\rm where}\quad x\equiv |{\bf x}|.
\label{hred}\end{equation}

Let us fix the constant energy surface
\begin{equation}
{\cal H}=E_{\rm osc}.\label{es}
\eeq
Then, dividing by $2\r^2 x$,
 we can represent it in the form
\beq
{\cal H}_{MIC}={\cal E},\quad
{\cal H}_{MIC}=\frac{(1+x)}{2\r^4}\left[{\bf p}^2
 + \frac{({\bf x} {\bf p})^2}{x}\right]
+ \frac{s^2}{2\r^4x^2} -\frac{\gamma}{\r^2 x}\quad,
\label{hr}\end{equation}
where  we introduced the notation
\beq
\gamma=E_{\rm osc}/2-s^2/\r^2,\quad  -2{\cal E}=\omega^2+s^2/\r^4 \quad.
\eeq
The Hamiltonian ${\cal H}_{MIC}$ can be  interpreted
as the Hamiltonian
of some generalized
MIC-Kepler problem.
Notice that  its potential energy term has the same form,
as the one of the
conventional (flat) MIC-Kepler problem.
The hidden symmetries of
 the system are given by the reduced generators $I_i$.

Let us perform the canonical
 transformation $({\bf x},{\bf p})\to({\bf{\tilde x}},
{\bf{\tilde p}})$,
 going  to the coordinates where the
metric takes a conformally-flat form:
\beq
{\bf {\tilde x}}= f(x){\bf x},\quad {\bf p}= {f} {\bf {\tilde  p}}
+f' \frac{({\bf x{\tilde p}})}{x}{{\bf x}}  , \label{conf}
\eeq
where
\beq
 f(x)=\frac 1x \frac{{\sqrt{1+x}}-1}{{\sqrt{1+x}}+1}\quad.
\eeq
In this case, the reduced Hamiltonian reads:
\beq
{\cal H}_{\rm red}=\frac{x(1+ x)^2{\bf{ p}^2}}{4\r^2}+
s^2\frac{(x+1)^4}{4\r^2 x(1-x)^2} +
\frac{4\omega^2\r^2 x}{(1-x)^2},
\quad\quad  x<1,
\eeq
while the Hamiltonian of the above obtained
generalization of MIC-Kepler problem
(\ref{hr}) takes the form
\beq
{\cal H}_{MIK}=\frac{(1- x^2)^2}{32\r^4}
\left( {\bf{ p}^2}+ \frac{s^2}{x^2}\right)
-(\gamma+\frac{s^2}{2\r^2} )\frac{1+x^2}{4\r^2x}-
\frac{s^2}{4\r^4}\quad.
\eeq
This  is nothing but the Hamiltonian of the  MIC-Kepler problem
on the three-dimensional hyperboloid \cite{np} constucted by the
 Kustaanheimo-Stiefel transformation of the oscillator on a
four-dimensional sphere.

Performing the Kustaanheimo-Stiefel transformation of the
system (\ref{hb}) on $\DC P^2$,
we  get the following expression for the
 reduced Hamiltonian:
\beq
{\cal H}_{Back}=\frac{(1+x)}{\r^2}\left[x{\bf p}^2 +
{({\bf x} {\bf p})^2}\right]
+ s^2\frac{1+x}{\r^2x} +
2\omega^2\r^2\frac{1+x}{x+x_3}-\omega^2\r^2,
 \quad x_3\neq x .
\eeq
In conformal  coordinates (\ref{conf}) the
latter  takes the form
\beq
{\cal H}_{\rm cBack}=\frac{x(1+ x)^2{\bf{ p}^2}}{4\r^2}+
s^2\frac{(x+1)^4}{4\r^2 x(x-1)^2} + {\omega^2\r^2}
\frac{(1+x)^2}{2(x+x_3)}- \omega^2\r^2.
\eeq

\setcounter{equation}0
\section{ ${\cal N}=2$ supersymmetric oscillator  on $\DC P^N$}
In this Section we construct the ${\cal N}=2$
superextension of the oscillator on
$\DC P^N$ coupled to a constant magnetic field.
It is well known that any Hamiltonian  system  of the form
\beq
{\cal H}_0=g^{ij}(p_i p_j+W_{,i}W_{,j}),
\quad\Omega^{\rm can}=dp_i\wedge dx^i
\label{vse}\eeq
could be easily extended to the system with exact
  ${\cal N}=2$ supersymmetry
 \begin{equation}
\begin{array}{c}
\{Q^+,Q^-\}={\cal H},\quad
\{Q^\pm ,Q^\pm \}=0.
\end{array}
\label{4sualg}\end{equation}
The function $W(x)$ is called superpotential.
The oscillator on a sphere $S^D$ belongs to the above
class of systems.
Its  superpotential is given by the expression
\beq
W=\frac{\omega}{2} \log\frac{2+{\bf x^2}}{2-{\bf x}^2} \quad,
\eeq
 where  ${\bf x}$ denotes the conformal
coordinates of the sphere $S^D$.

For the supersymmerization of the system (\ref{vse}),
 we have to  define the supersymplectic structure
$$
\Omega=dp_i\wedge dx^i +
 \frac 12 R_{ijkl}\theta^k_+\theta^l_- dx^i\wedge
dx^k +g_{ij}D\theta^i_+ \wedge D\theta^j_-,\quad
D\theta^i_\pm\equiv d\theta^i_\pm+\Gamma^i_{kl}\theta^k_\pm dx^l,
\quad \alpha=1,2
$$
and the supercharges $Q_{\pm}=(p_i\pm iW_{,i})\theta^i_\pm$,
 which obey the condition
$\{Q_\pm, Q_\pm \}=0$.
Then, we immediately get the  ${\cal N}=2$
supersymmetric Hamiltonian
$$
{\cal H}\equiv\{Q_+, Q_-\}={\cal H}_0 +W_{i;j}\theta^i_+\theta^j_-
+ R_{ijkl}\theta^i_-\theta^j_+\theta^k_-\theta^l_+.
$$
The inclusion of a magnetic field $\Omega\to
 \Omega+F_{ij}\theta^i_+\theta^j_-$
breaks the  ${\cal N}=2$ supersymmetry of the system
$$ \{Q_\pm,Q_{\pm}\}=F_{ij}\theta^i_\pm\theta^j_{\pm},
\quad    \{Q_+, Q_-\}=
{\cal H}+ iF_{ij}\theta^i_+\theta^j_- .$$

For the construction of the supersymmetric oscillator on
 $\DC P^N$, let us represent the initial (bosonic)
 Hamiltonian in the form
\beq
{\cal H}=g^{a\bar b}(\pi_a{\bar \pi}_b+
\partial_a W {\bar\partial}_b W).
\eeq
If the superpotential can be represented in the form
$W(z,\bar z)=W_+(z)+W_-(\bar z)$, then one can construct
 the ${\cal N}=4$ supergeneralization of the system  on K\"ahler
space \cite{Bellucci:2001ax}.
Otherwise, the system can be endowed
 with ${\cal N}=2$ supersymmetry.
Hence, we can construct the ${\cal N}=4$ supersymmetric
oscillator on $\DC^N$
choosing the superpotential $2W=\omega z^2+\omega {\bar z}^2$.
However,  we cannot  construct the (anti)holomorphic
 superpotential
 for the oscillator on $\DC P^N$ and, consequently,
obtain  its ${\cal N}=4$
superextension.
On the other hand, for the oscillators on $\DC^N$
and $\DC P^N$ one can find the
superpotentials with explicit $su(N)$ symmetry,
\beq
\begin{array}{ccc}
W=\omega K=\omega z\bar z&{\rm for }& \DC^N\\
2W=\omega\r\log (1-z\bar z)/(1+z\bar z)&{\rm for }&\DC P^1\\
W=\omega K=\omega\r\log (1+z\bar z)&{\rm for }&
 \DC P^N\;,\;N>1\quad .
\end{array}
\eeq
By using such functions,
 we shall construct the ${\cal N}=2$ supersymmetric oscillators
on $\DC P^N$.
We shall see that the linear dependence of the
superpotential  $W$ on the K\"ahler potential $K$
leads to an interesting behaviour  of the
supersymmetric system, with
 respect to a constant magnetic field.
Thus, the superoscillator  on $\DC P^N$, $N>1$
has more similarities with the
planar one, than the oscillator on $\DC P^1$.

Let us consider  a $(2N.2N)_{\DC}$-dimensional  phase space
equipped with the symplectic structure
\begin{equation}
\begin{array}{c}
\Omega=d\pi_a\wedge dz^a+ d{\bar\pi}_a\wedge d{\bar z}^a
+i(B g_{a\bar b}+iR_{a{\bar b}c\bar d}\eta^c_\alpha\bar\eta^d_\alpha)
dz^a\wedge d{\bar z}^b+
g_{a\bar b}D\eta^a_\alpha\wedge{D{\bar\eta}^b_\alpha}\quad,
\end{array}
\label{ss}\end{equation}
where $D\eta^a_\alpha
=d\eta^a_\alpha+\Gamma^a_{bc}\eta^a_\alpha dz^a,
\quad \alpha=1,2$, and
 $\Gamma^a_{bc},\; R_{a\bar b c\bar d}$ are,
respectively, the connection and curvature of
the K\"ahler structure.
The corresponding Poisson brackets are defined
by the following non-zero
relations (and their complex-conjugates):
$$
\begin{array}{c}
\{\pi_a, z^b\}=\delta^b_a,\quad
\{\pi_a,\eta^b_\alpha\}=-\Gamma^b_{ac}\eta^c_\alpha,\\
\{\pi_a,\bar\pi_b\}=i(Bg_{a\bar b}+
i R_{a\bar b c\bar d}\eta^c_\alpha{\bar\eta}^d_\alpha),
\quad
\{\eta^a_\alpha, \bar\eta^b_\beta\}=
g^{a\bar b}\delta_{\alpha\beta}.
\end{array}
$$
The symplectic structure (\ref{ss}) becomes canonical
in the coordinates $(p_a,\chi^k)$
\begin{equation}
\begin{array}{c}
p_a=\pi_a-\frac{i}{2} \partial_a{\bf g},
\quad\chi^m_i={\rm e}^m_b\eta^b_i:\quad
\Omega_{Scan}=dp_a\wedge d z^a +
d{\bar p}_{\bar a}\wedge d{\bar z}^{\bar a} +
iB g_{a\bar b}dz^a\wedge d{\bar z}^b
+d\chi^m_\alpha\wedge d{\bar\chi}^{\bar m}_\alpha,
\end{array}
\label{canonical}\end{equation}
where ${\rm e}^m_a$ are the einbeins of the
K\"ahler structure:
${\rm e}^m_a\delta_{m\bar m}{\bar{\rm e}}^{\bar m}_{\bar b}
=g_{a\bar b}.$

So, in order to quantize the system, one chooses
$$ {\hat p}_a=-i\left(\frac{\partial}{\partial z^a}-
iB\frac{\partial K}{\partial z^a}\right) ,\quad
 {\hat{\bar  p}}_{\bar a}=
-i\left(\frac{\partial}{\partial {\bar z}^{\bar a}}+i
B\frac{\partial K}{\partial {\bar z}^a}\right),\quad
[{\hat\chi}^m_\alpha,{\hat{\bar\chi}}^{\bar n}_\beta]_+
=\delta^{m\bar n}\delta_{\alpha\beta}.
$$
In order  to construct the system with the
exact  ${\cal N}=2$ supersymmetry
(\ref{4sualg},) we have to find  the appropriate
candidates for $Q^\pm$, which
obey the equations $\{Q^\pm,Q^\pm\}=0$.
Let us search the realization of
supercharges among the functions
\beq
Q^\pm=\cos\lambda\;\Theta^\pm_1 +\sin\lambda\;\Theta^\pm_2\;,
\eeq
where
\beq
\Theta^+_1=\pi_a \eta^a_1+ i\bar\partial_a W {\bar \eta}^a_2,\quad
\Theta^+_2={\bar\pi}_a\bar\eta^a_2 +i\;\partial_a W \eta^a_1,\quad
\quad
\Theta^-_{1,2}={\bar\Theta}^+_{1,2},
\eeq
and $\lambda$ is some parameter.
Calculating the Poisson brackets of the functions, we get
\begin{eqnarray}
&\{Q^+,Q^+\}=&
i(\sin 2\lambda\;Bg_{a\bar b}+
2\omega\cos 2\lambda W_{a\bar b})\eta^a_1\eta^b_2,\label{pp}\\
&  \{Q^+,Q^-\}=& {\cal H}^0_{SUSY}+
\cos 2\lambda\;B{\cal F}_3/2
-\sin 2\lambda\;{\cal Z}_3.
\label{pm}\end{eqnarray}
Here and in the following, we use the notation
\beq
{\cal H}^0_{SUSY}={\cal H}
-R_{a\bar b c\bar d}\eta^a_1\bar\eta^b_1\eta^c_2\bar\eta^d_2
-iW_{a;b}\eta^a_1\eta^b_2+
iW_{\bar a;\bar b}\bar\eta^a_1\bar\eta^b_2+ B
\frac{ig_{a\bar b}\eta^a_\alpha{\bar\eta}^b_\alpha }{2}, \label{hosup}
\eeq
where  ${\cal H}$ denotes the oscillator Hamiltonian
on $\DC P^N$ (see the expressions in (\ref{ho}), (\ref{hn})),
and
\beq
{\cal F}_3=
ig_{a\bar b}(\eta^a_1\bar\eta^b_1-\eta^a_2\bar\eta^b_2),
\quad
{\cal Z}_3=
i W_{a\bar b}(\eta^a_1\bar\eta^b_1-\eta^a_2\bar\eta^b_2),
%\quad
%{\bf g}=ig_{a\bar b}\eta^a_\alpha{\bar\eta}^b_\alpha \;.
\eeq
In what follows, we will also need the generators
\beq
{\cal F}_+=ig_{a\bar b}\eta^a_1\bar\eta^b_2,
\quad {\cal F}_-={\bar {\cal F}_+},
\eeq
which obey the commutation relations
\begin{eqnarray}
&\{{\cal F}_\pm, {\cal F}_3\}=
\mp 2i{\cal F}_\pm,\quad
\{{\cal F}_+, {\cal F}_-\}=i{\cal F}_3&\label{ff}\\
&\{{\Theta}^\pm_\alpha, {\cal F}_\pm\}=0,\quad
\{{\Theta}^\pm_\alpha, {\cal F}_\mp\}=
\pm i\epsilon_{\alpha\beta}{\Theta}^\mp_\beta,\quad
\{{\Theta}^\pm_\alpha, {\cal F}_3\}=
\pm i{\Theta}^\pm_\alpha,&\label{thf}
%\\
%&\{{\cal F}_\pm, {\bf g}\}=\{{\cal F}_3, {\bf g}\}=0,
%\quad
%\{{\Theta}^+_\alpha, {\bf g}\}=-i(\pi_a\eta^a_1-i\bar\partial_a W
%\bar\eta^a_2),\quad {\rm and}\;{\rm so}\;{\rm on}\; . &\label{other}
\end{eqnarray}
Comparing (\ref{pp}),(\ref{pm}) with (\ref{4sualg}),
 we can construct the
${\cal N}=2$ supersymmetric oscillator on
 $\DC P^N$.\\

{\bf\large Superoscillator on $\DC P^1$.}
Consider  the supersymmetrization of the
oscillator on the complex
projective plane $\DC P^1$.
Comparing  the equation (\ref{pp})
 with (\ref{4sualg}),
we get
\beq
\{Q^\pm, Q^\pm\}=0\Rightarrow \;
 B=0,\quad \cos 2\lambda=0,\quad \sin 2\lambda=\pm 1.
\eeq
Hence, we could choose {\it two}  copies of
 the supercharges and Hamiltonians
\beq
Q^{\pm}_\alpha=\frac{\Theta^{\pm}_1 -
(-1)^\alpha\Theta^{\pm}_2} {\sqrt{2}},\quad
\{Q^+_\alpha,Q^-_\alpha\}
\equiv {\cal H }_\alpha={\cal H}^0_{SUSY}
+(-1)^\alpha {\cal Z}_3,\quad \alpha=1,2 \;.
\eeq
{\it We  constructed two copies of the
 ${\cal N}=2$ supersymmetric
oscillator on $\DC P^1$.
The  inclusion of a constant magnetic
field $B$  breaks their  ${\cal N}=2$
supersymmetry  down to ${\cal N}=1$.}

Note that
$$\{Q^\pm_\alpha, Q^\pm_\beta\}=
2\epsilon_{\alpha\beta}{\cal Z}_\pm
\quad Z_\pm= A(z\bar z){\cal F}_\pm, \quad
Z_3=A(z\bar z){\cal F}_3,$$
where $A(z\bar z)=
\omega\frac{1+(z\bar z)^2}{(1-z\bar z)^2}$.
Hence, in the planar
 limit, one has $A\to\omega$, so that the generators
$Q^\pm_\alpha, Z_\pm, Z_3, {\cal H} $ form
 a closed Lie superalgebra.\\

{\bf\large{Superoscillator on $\DC P^N$, $N>1$.} }
On higher-dimensional complex projective
 spaces one has
\beq
W_{a\bar b}=\omega g_{a\bar b},\quad\Rightarrow
\;\{Q^\pm, Q^\pm\}=0\Leftrightarrow
B\sin\;2\lambda+2\omega\cos\;2\lambda=0.
\eeq
Let us introduce   the parameter  $\lambda_0$
\beq
\cos 2\lambda_0=\frac{B/2}{\sqrt{\omega^2+(B/2)^2}},\quad
\sin 2\lambda_0=-\frac{\omega}{\sqrt{\omega^2+(B/2)^2}},
\eeq
so that
\beq
\lambda=\lambda_0+(\alpha - 1)\pi/2,\quad \alpha=1,2.
\eeq
Hence, we get the following supercharges:
\beq
 Q^\pm_\alpha=\cos\lambda_0\Theta^\pm_1+
(-1)^\alpha\sin\lambda_0\Theta^\pm_2,
\eeq
and  the pair of corresponding ${\cal N}=2$
supersymmetric Hamiltonians
\beq
{\cal H}^\alpha_{SUSY}=
\{Q^+_\alpha,Q^-_\alpha\}=   {\cal H}^0_{SUSY} -(-1)^\alpha
{\sqrt{\omega^2+(B/2)^2}}{\cal F}_3.
\label{n4}\eeq

{\it We constructed,  on higher-dimensional
 complex projective spaces,
two copies of exact ${\cal N}=2$
 supersymmetric oscillators coupled to
a constant magnetic field.}

Calculating the commutators of $Q^\pm_1$
and  $Q^\pm_2$ we get
\beq
\{Q^\pm_1,Q^\pm_2\}=2\omega {\cal F}_\pm , \quad
\{Q^+_1,Q^-_2\}=0\,
\eeq
where the Poisson brackets between
${\cal F}_\pm $, and  $Q^\pm_\alpha$
look as follows:
\beq
\begin{array}{c}
\{Q^\pm_\alpha, {\cal F}_\pm\}=0,\quad
\{Q^\pm_\alpha, {\cal F}_\mp\}=
\pm\epsilon_{\alpha\beta}Q^\pm_\beta,\quad
\{Q^\pm_\alpha, {\cal F}_3\}=\pm iQ^\pm_\alpha.
\end{array}
\eeq
Hence, these two  systems form the superalgebra
\beq
\begin{array}{c}
\{Q^\pm_\alpha,Q^\pm_\beta\}=2\omega\epsilon_{\alpha\beta} {\cal F}_\pm,\quad
\{Q^\pm_\alpha,Q^\mp_\beta\}=
\delta_{\alpha\beta}{\cal H}^0_{SUSY}
-\sigma^3_{\alpha\beta}\omega^2 {\cal F}_3,\\
\{Q^\pm_\alpha, {\cal F}_\pm\}=0,\quad
\{Q^\pm_\alpha, {\cal F}_\mp\}=
\pm\epsilon_{\alpha\beta}Q^\pm_\beta,\quad
\{Q^\pm_\alpha, {\cal F}_3\}=\pm iQ^\pm_\alpha ,\\
\{{\cal F}_\pm,{\cal F}_\mp\}=i{\cal F}_3,
\quad\{{\cal F}_\pm,{\cal F}_3\}=\pm i{\cal F}_\pm \;.
\end{array}
\eeq
The symmetry superalgebra of the oscillator
on $\DC^N$ coincides with the
above one in any dimension, i.e., once again,
we find a quite different behaviour for
 the oscillators on $\DC P^1$ and
 $\DC P^N$, $N>1$  spaces, respectively.

Finally, we give the explicit expression
of the Noether constants
 of motion corresponding to
the $su(N)$ symmetry
 \beq
{\cal J}^{SUSY}_{a\bar b}={\cal J}_{a\bar b}+
\frac{\partial^2 {\ch}_{a\bar b}}{\partial z^c\partial
 {\bar z}^d}\eta^c\sigma_3\bar\eta^d .
 \eeq

\setcounter{equation}0
\section{Conclusion}
We  proposed an integrable system  on $\DC P^N$,
 with  $4N-1$
functionally independent
constants of motion, which could be viewed
  as the generalization of a
$2N-$dimensional oscillator.
 On the complex projective plane $\DC P^1=S^2$
this  system coincides
with the Higgs oscillator; the Kustaanheimo-Stiefel
transformation of the
system on $\DC P^2$
leads to the three-dimensional Coulomb-like system,
 which  is equivalent
to the MIC-Kepler problem on the
three-dimensional hyperboloid
obtained by the
 Kustaanheimo-Stiefel transformation
of the oscillator on $S^4$.
 On the other hand,
while the spherical oscillator remains unchanged
upon transition from one hemisphere to another,
  the oscillator on
$\DC P^N$, $N>1$, after transition to another chart,
yields a system which, in spite of the absence of
a rotational symmetry,
 remains ``maximally integrable".

The oscillators
on  $\DC P^3$ and $\DC P^4$, in our opinion,
deserve a separate study due to their relevance
to the higher-dimensional quantum
 Hall effect \cite{Zhang:2001xs}.
This theory, based on the quantum mechanics
of the particle on  $S^4$
interacting with a $SU(2)$ monopole field,
 lately has been
 extended to $\DC P^N$ spaces
in the presence of a constant
 $U(1)$ (magnetic) field \cite{kn}.
 Since $\DC P^3$ can be viewed as a
 fiber bundle of $S^4$ with $S^2$ in the bundle,
the four-dimensional quantum Hall system
 can be  formulated as a system
 on $\DC P^3$ \cite{kn,ber}.
 Performing the Hurwitz transformation
 of the oscillator on $\DC P^4$,
we will get the five-dimensional
 Coulomb-like system  with a $SU(2)$
Yang monopole.
 This system will have a degenerate ground state,
 hence it will be suitable for the developing of the five-dimensional
 quantum Hall effect in the Coulomb field
(in the present versions of higher-dimensional quantum Hall theory,
 the potential field is used
for the reduction to lower dimensions).

The K\"ahler structure makes the study of
the coupling of a constant magnetic field
to the  oscillator  on $\DC P^N$ much simpler than
on $2N$-dimensional sphere.
 In particular, we have shown that the
oscillators on $\DC^N$ and
 $\DC P^N$, $N>1$ coupled with a constant
magnetic field behave   similarly,
 with respect to ${\cal N}=2$
 supersymmetrization.
While a constant magnetic field breaks
the ${\cal N}=2$ supersymmetry
of the oscillator on sphere
 (and on the $\DC P^1=S^2$),
it preserves the ${\cal N}=2$
supersymmetry of the oscillators
on $\DC P^N$, $N>1$ and $\DC^N$.
On the other hand, in the absence of a
 magnetic field, the oscillator on $\DC^N$
allows us to introduce
${\cal N}=4$ supersymmetry,
while the oscillators
 on spheres and  $\DC P^N$  admit
only ${\cal N}=2$ superextensions.
It is easy to see, that the similarity
 of the oscillators
on $\DC^N$ and $\DC P^N$, $N>1$, in
their behaviour with respect
 to supersymmetrization,
is due to the special form of
 the Hamiltonian
$$
{\cal H}=g^{a\bar b}(\pi_a\bar\pi_b +
\omega^2\partial_a K \bar\partial_b K ),
$$
where $K$ is a K\"ahler potential of the metric.

Therefore, from the  viewpoint of ${\cal N}=2$ supersymmetry,
 the above Hamiltonian could be viewed  as
the generalization of the oscillator on an
 arbitrary K\"ahler manifold.
 In that case, the existence of hidden
symmetries of the oscillator on $\DC P^N$
 could be viewed as  an ``accidental" one.
 Simultaneously, it is clear that
 the oscillators on other
{\it symmetrical} K\"ahler spaces, say,
 on the  Lobachewski spaces ${\cal L}$,
 or Grassmanians ${\rm Gr}_{N.M}$,
will have  hidden symmetries, due to the
 translational
invariance of the above spaces.

\section*{{ Acknowledgments}}
We thank Erni Kalnins for stimulating
 questions that prompted us to undertake this
study, Pierre-Yves Casteill for checking the relations
(\ref{cpnalg}), (\ref{micalg}) on {\tt MATEMATICA}
  and  Anton Galajinsky  for  the interest in this work.
The work of S.B. was supported in part
by the European Community's Human Potential
Programme under
contract HPRN-CT-2000-00131 Quantum Spacetime,
the INTAS-00-0254 grant and the
NATO Collaborative Linkage Grant PST.CLG.979389.
The work of A.N was
 supported by grants INTAS 00-00262  and ANSEF  PS124-01.
A.N. thanks INFN-LNF
for hospitality during the completion of  this work.

\end{document}